\documentclass{article}
\usepackage{hiph-preprint}
\usepackage{graphicx}
\volnumber{22} \issuenumber{1} \edyear{2005}                             
\frompage{000} \topage{000}                                              
\recrevdate{1 January 2005}                                              
\def\rightmark{Quasiparticle picture of quarks near chiral transition}
\def\leftmark{M.~Kitazawa, T.~Kunihiro and Y.~Nemoto}

\newcommand{\bm}[1]{\mbox{\boldmath$#1$}}

\title{Quasiparticle picture of quarks near chiral phase transition} 
\authors{ 
{Masakiyo Kitazawa$^{1,2}$, Teiji Kunihiro$^1$, and Yukio Nemoto$^3$ %
\index{Kitazawa, M.} 
\index{Kunihiro, T.} 
\index{Nemoto, Y.}
}\\[2.812mm]
{\normalsize
\hspace*{-8pt}$^1$ 
Yukawa Institute for Theoretical Physics,\\
Kyoto University, Kyoto 606-8502, Japan\\[0.2ex] 
\hspace*{-8pt}$^2$
Institut f\"ur Theoretische Physik, J.W.\ Goethe-Universit\"at,\\
D-60054 Frankfurt am Main, Germany\\[0.2ex] 
\hspace*{-8pt}$^3$
Department of Physics, 
Nagoya University, Nagoya, 464-8602 Japan
}}
 
\abstract{
We study how the quasiparticle picture of the quark
can be modified near but above the critical tempearture ($T_c$)
of the chiral phase transition;
we incorporate into the quark self-energy
the effects of the precursory soft modes
of the phase transition, i.e., `para-$\sigma$($\pi$) meson'.
It is found that the quark spectrum 
has a three-peak structure near $T_c$:
We show that the additional new spectra originate from the 
mixing between a quark (anti-quark) and an anti-quark hole
(quark hole) caused by a ``resonant scattering'' 
of the quasi-fermions with the thermally-excited soft modes.
}

\makeindex
\begin{document}
 
\maketitle

\section{Introduction}

The properties of the quark-gluon plasma (QGP)
near the critical temperature $T_c$ of the QCD phase transition
have now attracted much interest.
The data from the RHIC experiment suggest that 
the hot matter near $T_c$ is a strongly coupled system.
The lattice simulations of QCD suggest the existence
of the hadronic bound state even above $T_c$.
Historically, the existence of the hadronic excitations 
in the QGP phase was first suggested 
for the light quark sector
as the soft modes of the chiral transition\cite{HK85}.

It has been shown recently\cite{KKKN04,KKN05a}
that the precursory {\em diquark} fluctuations\cite{KKKN02} 
are developed so as to form a soft mode 
at high density but at moderate temperature;
the soft mode in turn modifies the quark spectrum 
so significantly 
that there arises a depression of the quark spectrum around the
Fermi surface, leading to the formation 
of the {\em pseudogap} in the density of states.
It is thus highly expected that the precursory
soft modes of the chiral phase transition
should also strongly affect the quark spectrum near $T_c$.
In this report, 
we investigate how such soft modes 
affect the quark properties
 at $T\sim T_c$ but at vanishingly small chemical potential\cite{KKN05b}.

\section{Soft Modes and Quark Spectral Function}

To describe the quark matter near $T_c$,
we employ the two-flavor Nambu--Jona-Lasinio (NJL) model
in the chiral limit
\begin{eqnarray}
  \mathcal{L}=\bar{\psi} i \partial \hspace{-0.5em} / \psi
  + G_S [(\bar{\psi} \psi)^2 + (\bar{\psi}i\gamma_5\vec{\tau}\psi)^2],
\end{eqnarray}
with the coupling constant $G_S=5.5$ GeV${}^{-2}$ 
and the three dimensional cutoff $\Lambda=631$ MeV. 
This model gives the second order phase transition at $T_c=193.5$MeV
for vanishing quark chemical potential.

The soft modes of the chiral transition are characterized by 
the spectral function in the scalar and pseudo-scalar channels 
$\rho_{\sigma,\pi}(\bm{p},\omega) = -(1/\pi) D_\pi^R(\bm{p},\omega)$,
where $D_\sigma^R$ and $D_\pi^R$ are 
the quark-antiquark retarded Green functions in each channel\cite{HK85}.
One can show that 
there appear pronounced peaks in $\rho_{\sigma,\pi}(\bm{p},\omega)$
in the time-like region,
approximately around 
$\omega(\bm{p}) \simeq \pm\sqrt{ m^*_\sigma(T)^2 + |\bm{p}|^2 }$
with a $T$-dependent `mass'  $m^*_\sigma(T)$,
and as $T$ approaches $T_c$ the width of the peaks and $m^*_\sigma(T)$
become smaller\cite{HK85,KKN05b}. 

\begin{figure}[htb]
\begin{center}
\includegraphics[width=0.58\textwidth]{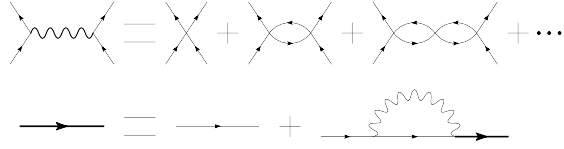}
\end{center}
\vspace*{-.5cm}
\caption[]{
The quark propagator in the RPA.
The wavy line denotes the soft modes of the chiral transition.
}
\label{fig:diagram}
\end{figure}

The existence of  the soft modes 
in turn  modifies the quark properties.
To incorporate the effect of the soft modes into the quark self-energy,
we employ the random phase approximation (RPA)
which is diagrammatically shown in Fig.~\ref{fig:diagram}.
The self-energy of the quark in the imaginary time formalism reads,
\begin{eqnarray}
  \tilde{\Sigma}(\bm{p},\omega_n) =
  T\sum_{m}\int\frac{d^3 q}{(2\pi)^3} 
  \left( {\cal D}_\sigma(\bm{p},\nu_n) +3{\cal D}_\pi(\bm{p},\nu_n) \right)
  \mathcal{G}_0(\bm{q},\omega_{m}),
\end{eqnarray}
where ${\cal D}_{\sigma,\pi}(\bm{p},\nu_n)$ are 
Matsubara propagator of the scalar and pseudo-scalar channels in the RPA
and ${\cal G}_0(\bm{p},\omega_n)$ denotes the free quark propagator.

The spectral function of the quark is expressed as
\begin{eqnarray}
{\cal A} (\bm{p},\omega) = 
-\frac1\pi {\rm Im} G^R (\bm{p},\omega)
= \rho_+(\bm{p},\omega) \Lambda_+(\bm{p})\gamma^0
+ \rho_-(\bm{p},\omega) \Lambda_-(\bm{p})\gamma^0,
\end{eqnarray}
with the retarded Green function of quark $G^R(\bm{p},\omega)$
and the projection operators 
$ \Lambda_\pm(\bm{q}) = (1\pm\gamma^0 \bm{\gamma}\cdot\bm{q}/|\bm{q}|)/2 $.

\section{Numerical Results and Discussions}

\begin{figure}[htb]
\vspace*{-.2cm}
\raisebox{7mm}{
\includegraphics[width=0.45\textwidth]{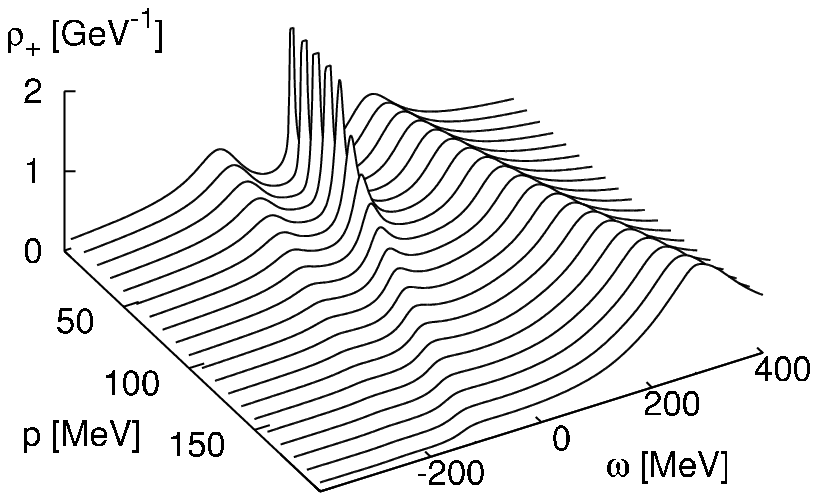}}
\includegraphics[width=0.37\textwidth]{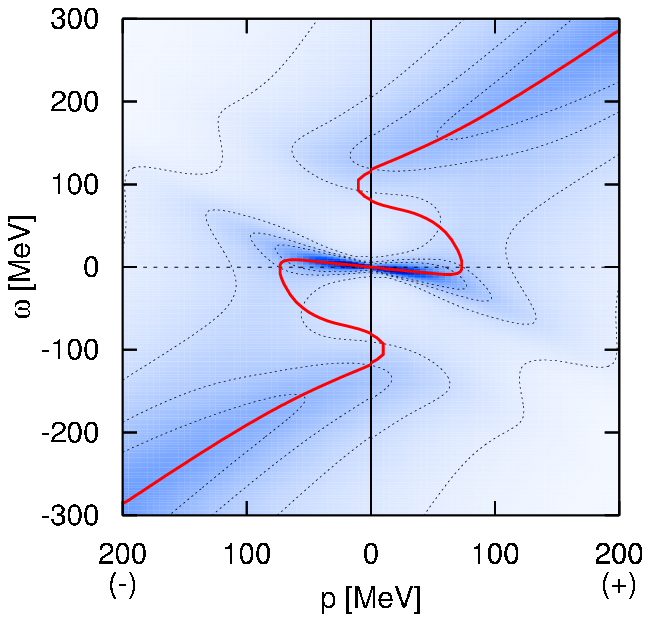}
\vspace*{-.2cm}
\caption[]{
The left panel shows the quark spectral function
$\rho_+(\bm{p},\omega)$ for 
$\varepsilon \equiv (T-T_c)/T_c = 0.1$.
The right panel shows
the contour map of $\rho_\pm( \bm{p},\omega )$ together with
the quasi-dispersion relations $ \omega=\omega_\pm(\bm{p}) $:
The right (left) half of the figure corresponds to
$\omega_+$ ($\omega_-$) and $\rho_+$ ($\rho_-$).
}
\label{fig:spectrum}
\end{figure}

In the left panel of Fig.~\ref{fig:spectrum},
we show the quark spectral function $\rho_+(\bm{p},\omega)$
for $\varepsilon\equiv (T-T_c)/T_c = 0.1$, i.e., slightly above $T_c$.
One sees a clear three-peak structure in the spectral function. 
We also show the contour map of $\rho_\pm(\bm{p},\omega)$
in the right panel
together with the quasi-dispersion relations $\omega_\pm(\bm{p})$
which are defined by zero of the real part of the inverse propagator.

\begin{figure}[htb]
\begin{center}
\includegraphics[width=0.53\textwidth]{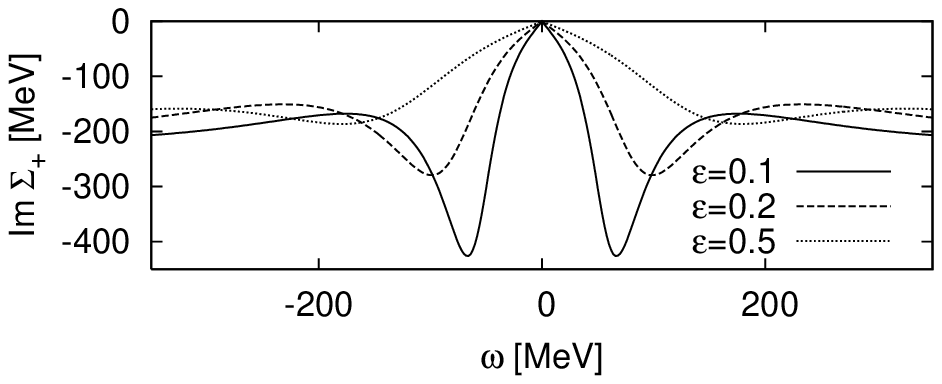}
\raisebox{10mm}{
\includegraphics[width=0.43\textwidth]{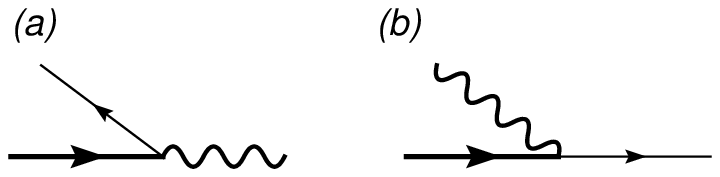}}
\end{center}
\vspace*{-5mm}
\caption[]{
Left panel:
The imaginary part of the quark self-energy
Im$\Sigma_+^R(\bm{0},\omega)$ for several values of 
$\varepsilon\equiv (T-T_c)/T_c$.
Right figures:
Parts of the physical processes which 
Im$\Sigma_+^R$ describes.
The thick lines represent the quasi-quarks,
the thin lines the on-shell free quarks 
and the wavy lines the soft mode.
}
\label{fig:sigma}
\end{figure}

In order to understand the mechanism of the appearance of the
three-peak structure in $\rho_\pm( \bm{p},\omega )$,
we show the imaginary part of $\Sigma_+^R( \bm{p}=0,\omega )$ 
in the left panel in Fig.~\ref{fig:sigma}.
One sees that
there develop two peaks in $|{\rm Im}\Sigma_+^R|$ 
in the positive- and negative-energy regions
at small temperatures,
with the peak positions moving 
toward the origin as $T$ is lowered to $T_c$.
It is found that 
the peaks in Im$\Sigma^R_\pm( \bm{p},\omega )$ in the positive and
negative energy regions essentially correspond to
the decay processes shown
in the right panel in Fig.~\ref{fig:sigma}(a) and (b), respectively;
 the wavy lines represent the soft modes.
One should notice here that the incident anti-quark line 
in Fig.~\ref{fig:sigma}(a) describe a thermally excited antiquark, 
which disappears after the collision with the soft modes. 
But the disappearance of the anti-quark means the
creation of a hole in the anti-quark distribution or a quark
number\cite{Weldon:1989ys}.
Figure~\ref{fig:sigma}(b) describes the decay process of a quasi-quark
which is a mixed state of quarks and antiquark-holes 
to an on-shell quark.
These processes induce a quark-`antiquark hole' mixing.

The mixing mechanism of quarks can be described 
in terms of the notion of {\em resonant scattering} 
as in the case of the (color-)superconductivity
\cite{JML97,KKN05a}, although a crucial difference
arises owing to the different nature of the soft modes.
In the case of the superconductivity,
the precursory soft mode 
is diffusion-mode like and has a strength around $\omega=0$.
The resonant scattering with the soft mode induces the mixing between
a particle and a hole, and gives rise to a gap-like
structure in the fermion spectrum around the Fermi energy; 
correspondingly, the imaginary part $|{\rm Im}\Sigma^R|$ of the 
quark self-energy has a single peak around 
the Fermi energy\cite{KKN05a}.
In the  present case,
the soft modes are propagating modes 
having a strength at finite energies
and hence the resonant scattering of the quarks with the
chiral soft modes gives rise to two peaks in
$|{\rm Im}\Sigma^R|$ at finite energies
and  induces a mixing between a quark (antiquark) and an
antiquark-hole (quark-hole).
Thus the two gap-like structures in the quark spectrum are formed
at the positive- and negative-energies.

In summary,
we have investigated the quark spectrum at $T$
near but above the critical temperature of the chiral phase transitions
taking account of the effects of the soft mode of the chiral transition.
We have shown that the quark spectrum has a three-peak structure 
at low momentum.
We have elucidated that the mechanism for realizing 
the three-peak structure can be understood as 
the mixing between a quark(anti-quark) and antiquark-hole (quark-hole)
induced by the {\em resonant scattering} of the quarks.

\vfill\eject

\begin{thebibliography}{99}  
  
\bibitem{HK85}
T.~Hatsuda and T.~Kunihiro,
Phys.\ Lett.\ B {\bf 145}, 7 (1984);
Phys.\ Rev.\ Lett.\  {\bf 55}, 158 (1985).

\bibitem{KKKN04}
M.~Kitazawa, T.~Koide, T.~Kunihiro and Y.~Nemoto,
Phys.\ Rev.\ D {\bf 70}, 056003 (2004);
Prog. Theor. Phys. {\bf 114}, 205 (2005).

\bibitem{KKN05a}
  M.~Kitazawa, T.~Kunihiro and Y.~Nemoto,
  arXiv:hep-ph/0505070, Phys. Lett. {\bf B}, in press.

\bibitem{KKKN02}
M.~Kitazawa, T.~Koide, T.~Kunihiro and Y.~Nemoto,
Phys.\ Rev.\ D {\bf 65}, 091504 (2002).

\bibitem{KKN05b}
  M.~Kitazawa, T.~Kunihiro and Y.~Nemoto,
  arXiv:hep-ph/0510167.

\bibitem{JML97}
B.~Jank\'o, J.~Maly and K.~Levin,
Phys. Rev. B {\bf 56}, R11407 (1997).

\bibitem{Weldon:1989ys}
H.~A.~Weldon,
%
Phys.\ Rev.\ D {\bf 61}, 036003 (2000).



\end{thebibliography}
\end{document}